\documentclass[12pt,a4paper]{iopart} 
\usepackage{amssymb,graphicx,cite}
\eqnobysec 
\begin{document}


 
 


 
 
   
 

\title[Fermionic dark solitons in a boson-fermion
mixture]{Free expansion of
fermionic dark solitons in a boson-fermion
mixture}
 
\author{Sadhan K Adhikari}
\address
{Instituto de F\'{\i}sica Te\'orica, Universidade Estadual
Paulista, 01.405-900 S\~ao Paulo, S\~ao Paulo, Brazil\\}

\date{\today}
 
 
\begin{abstract}

We use a time-dependent dynamical mean-field-hydrodynamic model to study
the formation of fermionic dark
solitons in a trapped degenerate fermi gas mixed with a Bose-Einstein
condensate in a harmonic as well as  a periodic 
optical-lattice potential.  The dark soliton with a ``notch"   in the
probability density with a zero at the minimum is simulated numerically as
a nonlinear continuation of the first vibrational excitation of the linear
mean-field-hydrodynamic equations, as suggested recently for pure bosons. 
 We  study the free expansion of these
dark solitons  as well as  the consequent increase in the size of 
their  central notch
and discuss the possibility of experimental observation 
of the notch after free expansion.

\end{abstract}

\maketitle

\section{Introduction}

Due to a strong repulsive Pauli-blocking interaction at low
energies among spin-polarized fermions,
there cannot be an  evaporative
cooling  leading to a quantum  degenerate fermi gas  (DFG)\cite{exp1}. 
Trapped DFG has
been achieved only by sympathetic cooling in the presence of a 
second
boson or fermion
component.  
Recently, there have been  successful observation
\cite{exp1,exp2,exp3,exp4} and associated  experimental
\cite{exp5,exp5x,exp6} and theoretical \cite{yyy,yyy1,zzz,capu,capu1,ska}
studies of
degenerate 
boson-fermion 
mixtures   by different experimental groups
\cite{exp1,exp2,exp3,exp4} in 
the following systems: $^{6,7}$Li \cite{exp3}, $^{23}$Na-$^6$Li
\cite{exp4} and 
$^{87}$Rb-$^{40}$K \cite{exp5,exp5x}. 
Also,  there have been 
studies of a degenerate mixture of two components of fermionic  
$^{40}$K \cite{exp1} and  
$^6$Li \cite{exp2} atoms. The
collapse of the  DFG in a boson-fermion mixture
$^{87}$Rb-$^{40}$K 
has been
observed and studied by Modugno {\it et al.} \cite{exp5,zzz,ska}. 
In these
studies of a mixed Bose-Einstein
condensate (BEC)
and a DFG  the initial
states 
were the stationary ground states of the systems.

In this paper 
we study the possibility of the formation of fermionic
dark solitons  in a mixture of a DFG with a BEC
using a coupled time-dependent mean-field-hydrodynamic
model where the  bosonic component is treated by the mean-field
Gross-Pitaevskii (GP)
equation \cite{11} and the fermionic component is treated by a
hydrodynamic
model \cite{capu,capu1}. This time-dependent mean-field-hydrodynamic model
was
suggested
recently by the present author \cite{ska} to study the collapse dynamics
of a DFG and 
is a
time-dependent extension of a time-independent model suggested for the
stationary  states by Capuzzi {\it et al.} \cite{capu,capu1} based
essentially 
on a Thomas-Fermi-Weizs\"acker approximation.

Zakharov and Shabat \cite{0b}
have shown that 
the 
dimensionless nonlinear Schr\"odinger 
(NLS) equation in the repulsive
or
self-defocusing case  \cite{1}
\begin{equation}\label{nls}
i u_t+u_{xx}-  |u|^2u=0.
\end{equation}
sustains the following
 dark and grey solitons \cite{5}:
\begin{eqnarray}\label{DS}
u(x,t)=r(x-ct) \exp[-i\{\phi(x-ct)-\mu t   \}],
\end{eqnarray}
with
\begin{eqnarray}r^2(x-ct)& = & \eta -2\kappa^2
\mbox{sech}^2[\kappa(x-ct)],  \\
\phi(x-ct)&=&\tan^{-1}[-2 \kappa/c \hskip 0.05cm \tanh\{\kappa (x-ct)\}],
\\
 \kappa &=& \sqrt{(2\eta - c^2)}/2,
\end{eqnarray}
where $c$ is the velocity of the soliton, $\mu$ the  parametric energy,
and $\eta$ related to intensity. Soliton (\ref{DS})
having  a ``notch" over a
background density is grey in general. It is dark if density
$|u|^2=0$ at the minimum.
The soliton can move freely with
velocity $c$ and at zero velocity the soliton becomes a dark soliton:
$|u(x,t)|= \sqrt {\eta} \tanh [x\sqrt{\eta/2}]$.  

The
similarity of the NLS equation (\ref{nls})
to the GP equation  (\ref{a}) (below)   
  imply  the possibility of a dark
soliton in a trapped
BEC \cite{bur}.  It has been suggested    that the dark soliton of a
trapped BEC
could be   a stationary eigenstate of the GP equation \cite{bur,5b} as in
the
case of the trap-less NLS equation. The usual search for a dark soliton in
the GP
equation proceeds through time evolution starting 
with the ansatz \cite{tanh,tanh1}
$u_{\small \mbox{DS}}= \tanh
(x) u_{\small \mbox{G}}(x)$, where $ u_{\small \mbox{G}}$ is the
ground
state
of the GP equation and the $\tanh (x)$ factor is introduced in analogy
with the dark soliton in the NLS equation.  However, the function
$u_{\small \mbox{DS}}$ is not an eigenfunction of the GP equation and
hence this procedure leads to numerical instability on time evolution
\cite{tanh,tanh1}. 
It has been demonstrated \cite{dsa} that 
the  time evolution of the GP equation
with the initial state  $u_{\small \mbox{DS}}$ 
leads to a dark soliton which is  the lowest vibrational excitation of the 
system. Exploring this, a stable  numerical procedure has been suggested
\cite{dsa}
for the simulation of dark soliton which we use in this investigation.

We study 
the formation of fermionic dark solitons  in a DFG mixed
with a BEC in a  harmonic as well as   a periodic  optical-lattice trap. 
There have been  experimental \cite{exdks} and theoretical \cite{thdks}
studies of the formation of   dark solitons in a
harmonically
trapped BEC. In view of this, here we study   for the first time 
the possibility of the
formation of a fermionic dark soliton in a  DFG using a mean-field model
for a mixture of DFG and BEC.

Collective excitations in the form of solitons and vortices in trapped
fermions have also been investigated recently by Damski et al. \cite{x}
and Karpiuk et al. \cite{y}. However, they considered isolated ultra-cold
fermions and not a realistic mixture of trapped DFG and BEC as in the
experiments and as discussed in this paper.  Also they did not demonstrate
the existence of stable dark solitons with a zero at the center of the
notch as in the present study. They identified grey soliton-like structure
with a shallow dip in the isolated fermionic density distribution quite
distinct from the stable fermionic dark solitons in a DFG-BEC mixture as
noted in this investigation.

Though the present dark solitons are numerically stable in the
mean-field formulation, they could be  unstable physically 
due to quantum
fluctuations
\cite{z}. The effect of quantum fluctuations is lost in the mean-free
model and can only be studied in a field-theoretic approach.
Moreover being an excited state they are thermodynamically unstable.  
There have been suggestions about how to excite a dark soliton by phase
imprinting
method \cite{x,y}. 
The dark soliton is the lowest vibrational excitation of the 
BEC \cite{dsa} and there have also been investigations about how to attain
such
excited
states \cite{yu}. 
Nevertheless, despite different   
suggestions about how to excite a dark soliton  \cite{x,y,yu}, experiments
to date have not yet generated a stationary
dark soliton. 
Experimentally, so far the dark solitons have been unstable 
\cite{bur,5b}.  
Although we cannot make definite suggestion(s) for the formation of
stable fermionic dark solitons, considering that they are stationary
excitations
of the mean-field equation their creation seems possible at least as a
non-stationary dark soliton which may turn grey and oscillate before
decaying due to quantum fluctuations and thermodynamic effects.

Lately, the periodic optical-lattice potential has played an essential
role in many theoretical and experimental studies of  Bose-Einstein
condensation, e. g., in the study of Josephson oscillation \cite{j} and
its disruption \cite{jd},
interference
of matter-wave \cite{i}, BEC dynamics on periodic trap \cite{d}, etc. The
periodic optical-lattice confinement generated experimentally by a
standing-wave laser
field  
creates a BEC in an entirely different
shape and trapping condition form a conventional harmonic oscillator
trapping. In view of this we study the possibility of the formation of a
fermionic dark soliton in an optical-lattice potential. The formation of a
bosonic dark soliton in an optical-lattice potential has already been
investigated \cite{tanh1}.

The central notch is the earmark of a dark soliton.  Experimentally, a
dark soliton is identified after removing the traps so that a free
expansion of the DFG allows the notch to widen and be photographed
clearly.  In view of this we study a free expansion of dark solitons in a
DFG-BEC mixture and study the possibility of detection of a fermionic dark
soliton in the laboratory.

In section 2 we present an account of the time-dependent mean-field model
consisting of a set of coupled partial differential equations 
involving a BEC and a DFG.
In the case
of a cigar-shaped system with stronger  radial trapping, 
the above model is reduced to an effective
one-dimensional form appropriate for the study of dark solitons. In
section 3   we present our results for stationary fermionic dark
solitons as well as a
study of their  free expansion in a boson-fermion mixture. 
Finally,
a summary of our findings  is given in section 4.
 
\section{Nonlinear mean-field-hydrodynamic model}

The time-dependent BEC wave
function $\Psi({\bf r},t)$ at position ${\bf r}$ and time $t $
may
be described by the following  mean-field nonlinear GP equation
\cite{11}
\begin{eqnarray}\label{a} \biggr[- i\hbar\frac{\partial
}{\partial t}
-\frac{\hbar^2\nabla_{\bf r}^2   }{2m_{{B}}}
+ V_{{B}}({\bf r})
+ g_{{BB}} n_B
 \biggr]\Psi_{{B}}({\bf r},t)=0, 
\end{eqnarray}
with normalization $ \int d{\bf r} |\Psi_B({\bf r},t)|^2 = N_B. $ 
Here $m_{{B}}$
is
the mass and  $N_{{B}}$ the number of bosonic atoms in the
condensate, $n_B\equiv  |\Psi_{{B}}({\bf r},t)|^2$ is the boson 
probability density,
 $g_{{BB}}=4\pi \hbar^2 a_{{BB}}/m_{{B}} $ the strength of
inter-atomic interaction, with
$a_{{BB}}$ the boson-boson scattering length. 
The trap potential with axial symmetry may be written as  $
V_{{B}}({\bf
r}) =\frac{1}{2}m_B \omega ^2 (\rho^2+\nu^2 z^2)$ where
 $\omega$ and $\nu \omega$ are the angular frequencies in the radial
($\rho$) and axial ($z$) directions with $\nu$ the anisotropy parameter.
The probability density   $n_F$ of an isolated DFG  in the 
Thomas-Fermi
approximation 
is given by
\cite{zzz}
\begin{eqnarray}\label{b}
n_F= \frac{[\mbox{max}(0,\{\epsilon_F-V_F({\bf r})\})]^{3/2}}{A^{3/2}},
\end{eqnarray}
where 
$A=\hbar^2 (6
\pi^2
)^{2/3}/ (2m_F)$, $\epsilon_F$ is the Fermi energy, $m_F$ is the
fermionic mass, and the function $\mbox{max}$  denotes the larger of the
arguments. The confining trap potential
$V_F({\bf
r})$ has axial symmetry. The number of fermionic atoms $N_F$
is given by the normalization $\int d{\bf r} n_F=N_F$.

We developed a set of practical time-dependent mean-field-hydrodynamic
equations for the interacting boson-fermion mixture starting from the
following Lagrangian density \cite{ska} \begin{eqnarray}\label{yy} {\cal
L}&=& \frac{i}{2}\hbar \left[ \Psi_B\frac{\partial \Psi_B^*}{\partial t} -
\Psi_B^* \frac{\partial \Psi_B}{\partial t} \right] +
\frac{i}{2}\hbar \left[ \sqrt{n_F}\frac{\partial {\sqrt n_F} ^*}{\partial
t} - {\sqrt n_F}^* \frac{\partial \sqrt{n_F}}{\partial t} \right]
\nonumber \\ &+& \left(\frac{\hbar^2|\nabla_{\bf r} \Psi_B|^2
}{2m_B}+V_B|\Psi_B|^2+\frac{1}{2}g_{BB} |\Psi_B|^4\right)\nonumber \\ &+&
\left(\frac{\hbar^2 |\nabla_{\bf r} \sqrt{n_F}|^2 }{6m_F}+
V_F|n_F|+\frac{3}{5} A |n_F|^{5/3}\right)\nonumber \\ &+& g_{BF} n_F
|\Psi_B|^2, \end{eqnarray} where $g_{BF}=2\pi \hbar^2 a_{BF}/m_R$ with the
boson-fermion reduced mass $m_R=m_Bm_F/(m_B+m_F),$ where $ a_{BF}$ is the
boson-fermion scattering length.

It
may not be entirely proper to define an average fermionic
wave function  $\Psi_F=\sqrt n_F$ in a DFG
like in a BEC. The correct fermionic wave function is to be
calculated from a
Slater determinant Schr\"odinger equation for degenerate fermions 
\cite{yyy1}.   
However, the probability
density $n_F$ of a DFG calculated in this fashion should lead to
reasonable results
\cite{yyy1} and has led to proper probability distribution 
for a DFG
\cite{capu1,ska} as well as results for collapse of a 
DFG \cite{zzz,capu1,ska} in agreement with experiment. This approach has
also been used successfully to predict a fermionic bright soliton in a 
boson-fermion mixture \cite{fbs}.

The terms in the first round bracket   on the right-hand side of
 (\ref{yy}) are the standard 
Gross-Pitaevskii terms  for the bosons and are related to a
Schr\"odinger-like equation \cite{11}. However, 
terms in the second  round bracket, although bear
a resemblance with the first,
are derived from the hydrodynamic equation of motion of the fermions
including a Weisz\"acker kinetic energy 
and are not related to a  Schr\"odinger-like equation \cite{capu}. Hence,
the second kinetic energy term 
has a different mass factor $6m_F$ and not the
conventional  Schr\"odinger mass factor $2m_B$ as in the first term.
Finally, the 
last term in this equation 
corresponds to an interaction between bosons
and fermions. The interaction between bosons and between
bosons and fermions are described by contact potentials parametrized by
coupling constants $g_{BB}$ and $g_{BF}$ defined above.
The interaction between fermions in
spin polarized state is highly suppressed due 
to Pauli blocking 
and has been neglected in  (\ref{yy})  and will be
neglected throughout this paper.

Recently, Jezek {\it et al.} \cite{jz} used the Thomas-Fermi-Weizs\"acker
kinetic energy term $T_F$ of fermions in their formulation which, in our
notation, will correspond to a fermionic kinetic energy of
$\hbar^2|\nabla_{\bf r}\sqrt{n_F}|^2/(9m_F) $ in  (\ref{yy}) in place
of the present term $\hbar^2|\nabla_{\bf r}\sqrt{n_F}|^2/(6m_F) $.  This
kinetic energy term contributes little to this problem compared to the
dominating $3A|n_F|^{5/3}/5$ term in  (\ref{yy}) and is usually
neglected in the Thomas-Fermi approximation.  However, its inclusion leads
to a probability density  which is a smooth and analytic function of the
space variable  \cite{jz}. For a
discussion of these two fermionic kinetic energy terms we refer the reader
to  \cite{capu,jz,pi}.

With the Lagrangian density (\ref{yy}), the Euler-Lagrange equations of
motion become \cite{ska}:  \begin{eqnarray}\label{e} \biggr[ &-&
i\hbar\frac{\partial }{\partial t} -\frac{\hbar^2\nabla_{\bf
r}^2}{2m_{{B}}} + V_{{B}}({\bf r}) + g_{{BB}}n_B + g_{{BF}}
|n_F|
 \biggr]\Psi_{{B}}({\bf r},t)=0,
\end{eqnarray}
\begin{eqnarray}\label{f} \biggr[& -& i\hbar\frac{\partial
}{\partial t}
-\frac{\hbar^2\nabla_{\bf r}^2}{6m_{{F}}}
+ V_{{F}}({\bf r}) 
+ A |n_F|^{2/3} 
+ g_{{BF}} n_B
 \biggr]\sqrt{n_{{F}}({\bf r},t)}=0. 
\end{eqnarray}

When the nonlinearity in   (\ref{f})
is  very large, 
the kinetic energy term in this equation can be neglected and the
time-independent stationary form of this equation becomes
\begin{equation}\label{mod}
n_F= \frac{[\mbox{max}(0,\{\epsilon_F-V_F({\bf
r})-g_{BF}n_B\})]^{3/2}}{A^{3/2}},
\end{equation}
which is the generalization of  (\ref{b}) in the presence of
boson-fermion coupling. Equation (\ref{mod}) has been used by 
Modugno
{\it et al.} \cite{zzz} for an analysis of a BEC coupled to  a DFG. We
shall see in the following that in actual experimental
condition the nonlinearity in   (\ref{f})    is quite large and  
(\ref{mod}) is a good approximation. We note that the
Lagrangian density of the formulation of Jezek {\it et al.} 
\cite{jz} reduces to the present Lagrangian density in this
approximation upon the neglect of the fermionic kinetic energy term.

The solution of the coupled three-dimensional equations above 
for studying  dark solitons in a
boson-fermion mixture is
a formidable task. Hence, 
we shall reduce  (\ref{e}) and  (\ref{f}) to the minimal 
one-dimensional form suitable 
for the study of dark solitons in a cigar-shaped geometry for $\nu <<1$. 
We perform this
reduction below where we 
take $V_B({\bf r})=V_F({\bf r})= \frac{1}{2}m_B\omega^2(
\rho^2+\nu^2 z^2)+V_0\sin^2(2\pi z/ \lambda )$ which corresponds to a
suppression  of $\omega$ and
$\nu  \omega$ for fermions 
by a factor
$\sqrt{m_B/m_F}$ as in the study by 
Modugno {\it et al.} \cite{zzz} and Jezek  {\it et al.} \cite{jz}. 
In the confining potential we also include the following optical-lattice
potential: $V_0\sin^2(2\pi z/ \lambda )$ \cite{catas}. Here $V_0$ is the
strength of
the optical-lattice potential and $\lambda $ is the wave length of the
laser.

For  $\nu << 1$,  (\ref{e})
and (\ref{f}) can be reduced to an effective 
one-dimensional form by considering solutions of the type
$\Psi_B({\bf r},t)=  \phi_B(z,t)\psi_B^{(0)}( \rho)$ and 
$\sqrt{n_F({\bf r},t)}=  \phi_F(z,t)\psi_F^{(0)}( \rho), $ 
where 
\begin{eqnarray}
|\psi_i^{(0)}(\rho)|^2&\equiv&
{\frac{M_i\omega}{\pi\hbar}}\exp\left(-\frac{M_i
\omega 
\rho^2}{\hbar}\right), \quad i=B,F,
\end{eqnarray}
corresponds to the respective circularly symmetric 
ground state wave function in the absence of nonlinear interactions and
satisfies
\begin{eqnarray}
-\frac{\hbar^2}{2m_B}\nabla_\rho ^2\psi_B^{(0)}
+
\frac{1}{2}m_B\omega^2\rho^2
\psi_B^{(0)}&=&\hbar\omega
\psi_B^{(0)},\\
-\frac{\hbar^2}{6m_F}\nabla_\rho^2\psi_F^{(0)}+
\frac{1}{2}m_B\omega^2\rho 
^2\psi_F^{(0)}&=&\sqrt{\frac{m_B}{3m_F}}\hbar\omega
\psi_F^{(0)},\nonumber \\
\end{eqnarray}
with normalization 
$2\pi \int_{0}^\infty |\psi_i^{(0)}(\rho)|^2 \rho d\rho=1.$
Now the dynamics is carried by $ \phi_i(z,t)$ and the radial dependence is
frozen in the ground state $\psi_i^{(0)}(\rho)$. The separation of
the variables is suggested by the structure of  (\ref{e}) and
(\ref{f}).

Averaging over the radial mode $\psi_i^{(0)}(\rho)$, 
i.e., multiplying
 (\ref{e}) and (\ref{f})
by  $\psi_i^{(0)*}(\rho)$
and integrating over $\rho$, we obtain the following one-dimensional 
dynamical equations \cite{abdul}:
\begin{eqnarray}\label{i} \biggr[ &-& i\hbar\frac{\partial
}{\partial t}
-\frac{\hbar^2}{2m_{{B}}}\frac{\partial^2}{\partial z^2}
+ \frac{1}{2}m_B\nu^2 \omega^2 z^2
+V_0\sin^2\left( \frac{2\pi
z}{\lambda} \right)   \nonumber \\  &+&F_{BB}|
\phi_B|^2
+ F_{BF}| \phi_F|^2
 \biggr] \phi_{{B}}(z,t)=0, 
\end{eqnarray}
\begin{eqnarray}\label{j} 
\biggr[& -& i\hbar\frac{\partial
}{\partial t}
-\frac{\hbar^2}{6m_F}\frac{\partial^2}{\partial z^2}
+ \frac{1}{2}m_B\nu ^2\omega^2 z^2+V_0\sin^2\left( \frac{2\pi
z}{\lambda}\right) \nonumber \\  
&+&F_{FF}|
\phi_F|^{4/3}  
+ F_{BF}| \phi_B|^2
 \biggr] \phi_{{F}}(z,t)=0, 
\end{eqnarray}
where 
\begin{eqnarray}
 F_{BB}=g_{BB}\frac{\int_0^\infty|\psi_B^{(0)}|^4\rho d\rho}
{\int_0^\infty|\psi_B^{(0)}|^2\rho d\rho}=
g_{BB}{\frac{m_B\omega}{2\pi\hbar}},
\end{eqnarray}
\begin{eqnarray}
F_{BF}=g_{BF}\frac{\int_0^\infty|\psi_F^{(0)}|^2|\psi_B^{(0)}|^2\rho 
d\rho}{\int_0^\infty|\psi_B^{(0)}|^2\rho d\rho}
=g_{BF}{\frac{M_{BF}\omega}{\pi\hbar}},
\end{eqnarray}
\begin{eqnarray}
F_{FF}=A
\frac{\int_0^\infty|\psi_F^{(0)}|^{2+4/3}\rho
d\rho}{\int_0^\infty|\psi_B^{(0)}|^2\rho
d\rho} =
{\frac{3A}{5}}\left[
\frac{M_F\omega}{\pi \hbar}    \right]^{2/3}.
\end{eqnarray}
In  (\ref{i}) and (\ref{j}) we have included the optical-lattice 
potential and 
the normalization there 
is given by $\int_{-\infty}^\infty |\phi_i(z,t)|^2
dz = N_i$.

For calculational purpose it is convenient to reduce 
the sets   (\ref{i}) and (\ref{j})  to
dimensionless form 
by introducing convenient  dimensionless variables. Although the algebra is
quite straightforward, the expressions become messy with  different
factors
of masses. 
As we shall not be interested in a particular
boson-fermion 
system in this
paper, but  be concerned with the formation of fermionic
dark solitons 
in general,  we take in the rest of this paper $m_B=3 m_F=
m(^{87} \mbox{Rb})$, 
whence $m_R=3m_F/4, M_B=M_F=m_B,$ and
$M_{BF}=m_B/2 $ and where $m(^{87}{\mbox{Rb}})$ is the mass of the Rb
atom.
In
the two experimental
situations of \cite{exp4,exp5} 
$m_B \approx 
3m_F$.

In  (\ref{i}) and (\ref{j}), $\nu << 1 $ and  
we consider the dimensionless variables 
$\tau=t\nu \omega/2$,
$y=z /l_z$,
${\chi}_i=
\sqrt{(l_z/N_i)} \phi_i$, with $l_z=\sqrt{\hbar/(\nu \omega m_B)}$, 
so that 
\begin{eqnarray}\label{m} \biggr[& - & i\frac{\partial
}{\partial \tau}
-\frac{d^2}{dy^2} 
+ y^2+v_0\sin^2\left(\frac{2 \pi y}{\lambda_ 0}   
  \right)+    
 N_{BB}
\left|{{\chi}_B}\right|^2\nonumber \\
&+& N_{BF}
  \left|{{\chi}_F}\right|^2                  
 \biggr]{\chi}_{{B}}({y},\tau)=0,         
\end{eqnarray}
\begin{eqnarray}\label{n} \biggr[& - & i\frac{\partial
}{\partial \tau}-\frac{d^2}{dy^2} 
+ y^2+v_0\sin^2\left(\frac{2 \pi y}{\lambda_ 0}     \right)+ 
N_{FB}
  \left|{{\chi}_B} \right|^2 \nonumber \\
&+&
N_{FF}
  \left|{{\chi}_F}
\right|^{4/3}
 \biggr]{\chi}_{{F}}(y,\tau)=0,
\end{eqnarray}
where
$N_{BB}=(4/\nu)a_{BB}N_B/l_z,$
$N_{BF}=(8/\nu)a_{BF}N_F/l_z,$ 
$N_{FB}=(8/\nu)a_{BF}N_B/l_z,$ and
$N_{FF}=9(6\pi N_F/\nu)^{2/3}/5. $
 In 
(\ref{m}) and (\ref{n}),
the normalization condition  is given by 
$\int_{-\infty}^\infty |\chi_i(y,\tau)|^2 dy =1 $ and 
$v_0\equiv 2V_0/(\hbar \omega\nu)$ is the reduced strength of the
optical-lattice potential and $\lambda_0\equiv \lambda /l_z$ is the
dimensionless wave length.

\section{Numerical Result}

The coupled mean-field-hydrodynamic  equations 
 (\ref{m}) and
(\ref{n}) for dark solitons are solved 
numerically using a time-iteration
method based on the Crank-Nicholson discretization scheme
elaborated in  \cite{sk1}.  
We
discretize the mean-field-hydrodynamic  equations
using time step $0.0005$ and space step $0.025$.

We performed the time evolution of the set of equations (\ref{m}) and
(\ref{n})  setting  
$N_{BB}=N_{BF}=N_{FB}=N_{FF}=v_0=0$  and starting with the
eigenfunction of
the lowest excited state of the 
linear harmonic oscillator problem as suggested recently \cite{dsa},
e.g., with
$\chi_B(y,\tau)=\chi_F(y,\tau)=
\sqrt 2 \pi^{-1/4}y\exp(-y^2/2)\exp(-3i\tau)$. 
During the course of time evolution the nonlinear terms are switched on
very slowly and the resultant solutions iterated  (about 50000 times)
until convergence was
obtained.   If converged solutions are obtained, they correspond to the
required dark solitons.  In the present approach the time evolution starts
with and proceeds through successive eigenfunctions of the coupled
mean-field equations (\ref{m}) and (\ref{n}). Hence it leads to stable
numerical results. The usual numerical procedure for the calculation of
the dark solitons starts with an approximate solution 
of the mean-field
equation and hence leads to numerical instability on time evolution
\cite{tanh,tanh1,dsa}.

\begin{figure}
 
\begin{center}
\includegraphics[width=0.6\linewidth]{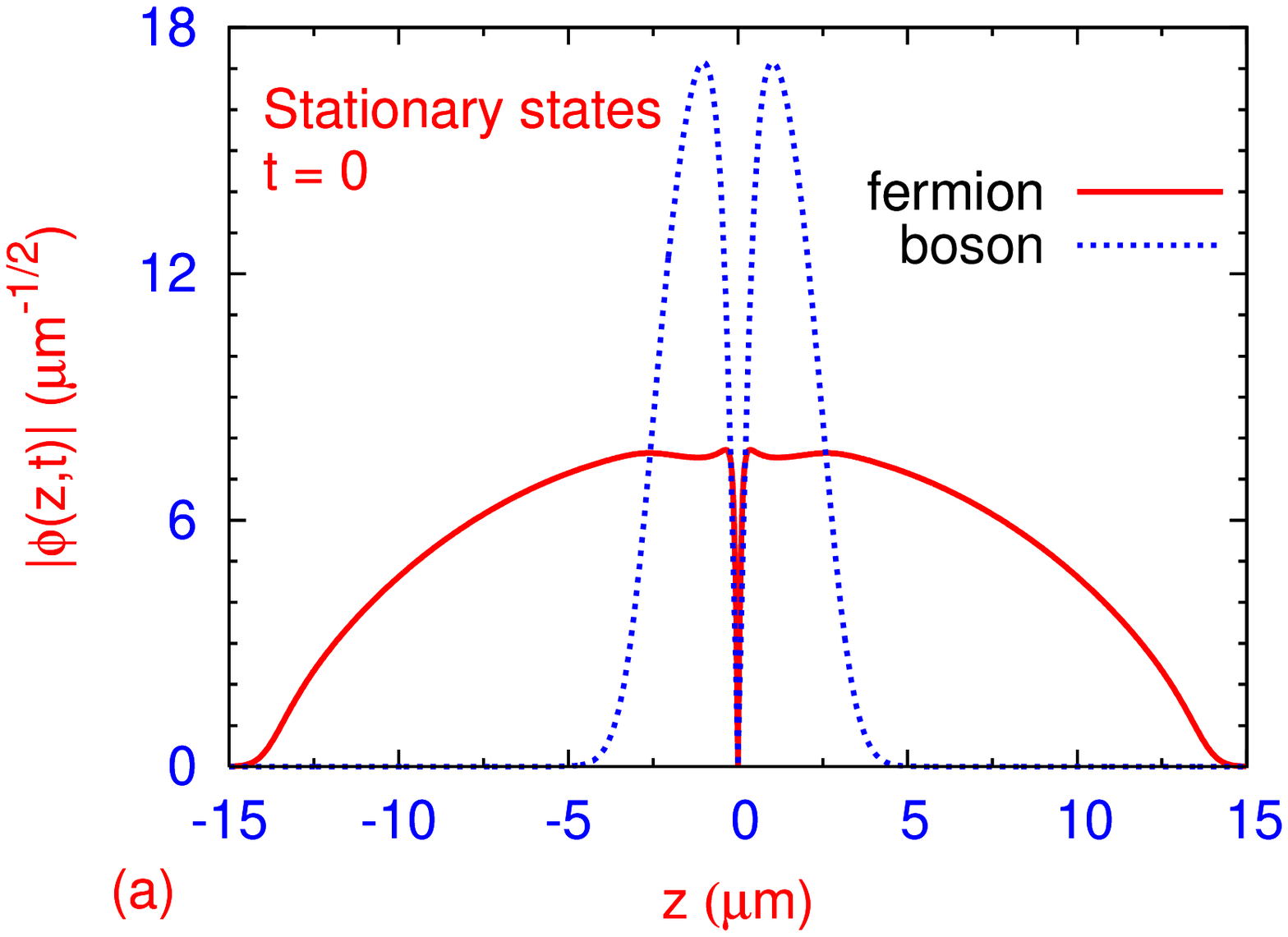}
\includegraphics[width=0.6\linewidth]{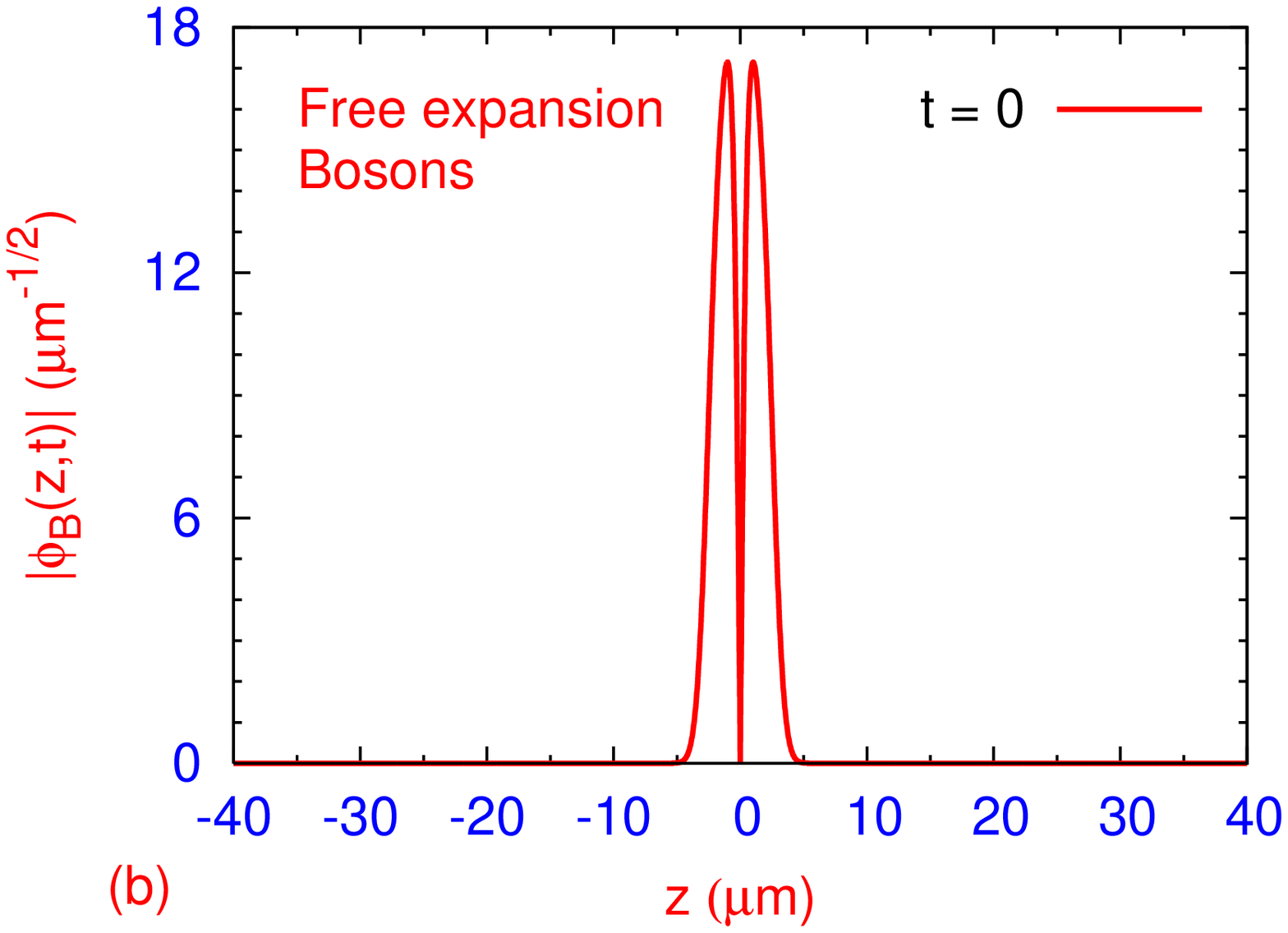}
\includegraphics[width=0.6\linewidth]{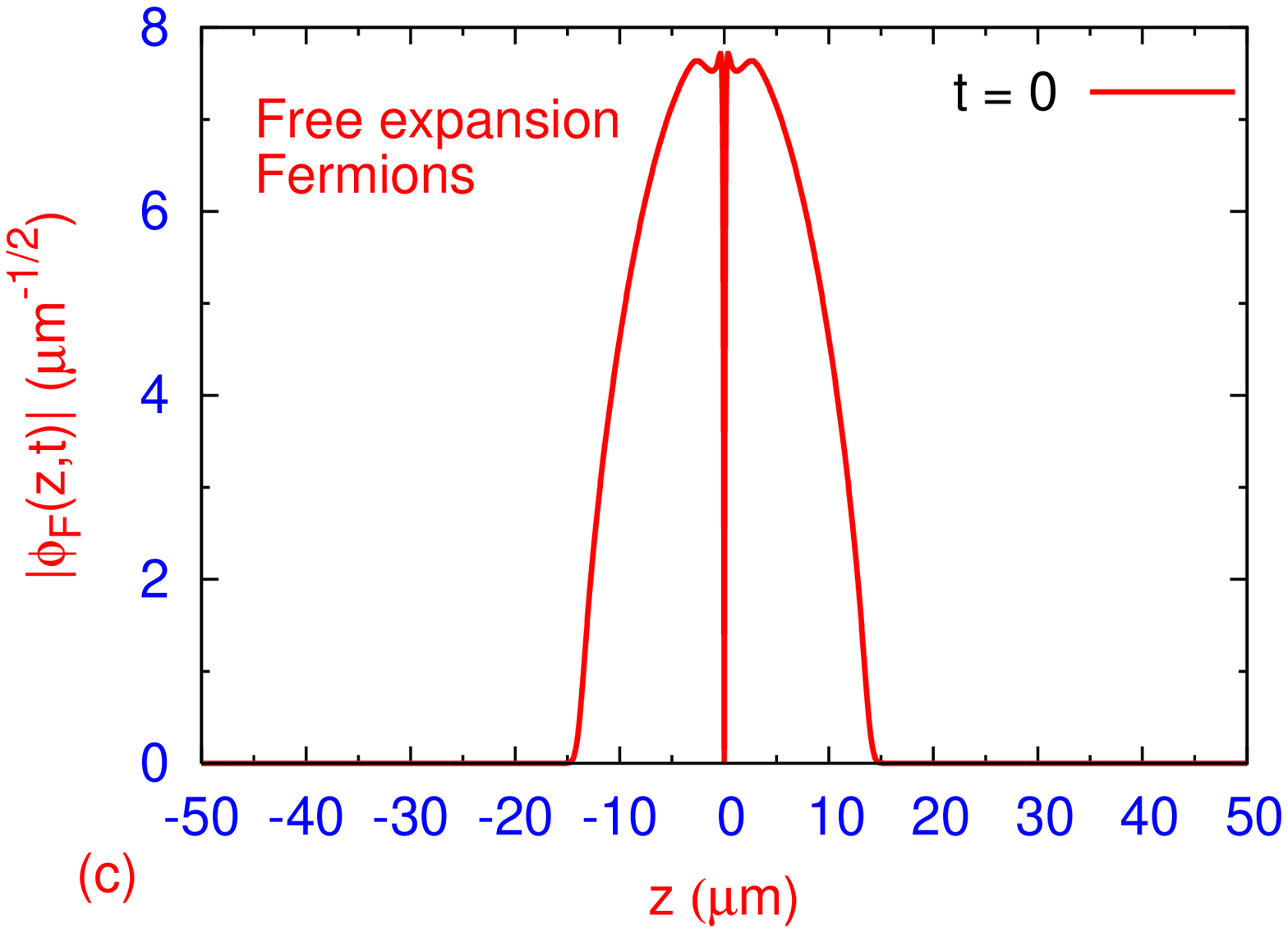}
\end{center}

\caption{(a) The
stationary function $|\phi(z,t)|$ for bosonic and
fermionic dark solitons  vs. $z$ with  $N_F=N_B=100$, $\nu=0.1$
 and $a_{BB}=a_{BF}=5 $ nm.  We show in
(b) and (c) the profiles of the bosonic and fermionic 
functions  $|\phi_B(z,t)|$ and  $|\phi_F(z,t)|$, respectively, of the
degenerate mixture (a) during
free expansion at regular intervals of
time. The nonlinearities are $N_{BB}=20, N_{BF}=40, N_{FB}=40,
N_{FF}=1275.$}

\end{figure}

We solve  (\ref{m}) and (\ref{n}) for dark solitons. 
In this case  the nonlinearity $N_{FF}$ could be very large for 
$N_F>100$, which may require special care 
for obtaining  accurate numerical 
solutions. In our calculation we use $\nu = 0.1$, $v_0=0$,
 $\nu \omega= 2\pi \times 100$ Hz and $m_B$ to be the mass of
$^{87}$Rb. Consequently, $l \approx 1$ $\mu$m and unit of time
$\tau=2/(\nu
\omega)$ is 3 ms. We also take  $N_F=100$, 
$N_B=100$, $l_z=1$ $\mu$m, and $a_{BB}=a_{BF}=5$ nm. 
With
these
parameters the nonlinearities in  (\ref{m}) and (\ref{n}) are 
$N_{BB}=20 $, $N_{BF}= 40$, $N_{FB}= 40$, and $N_{FF}= 1275$.

\begin{figure}
 
\begin{center}
\includegraphics[width=0.6\linewidth]{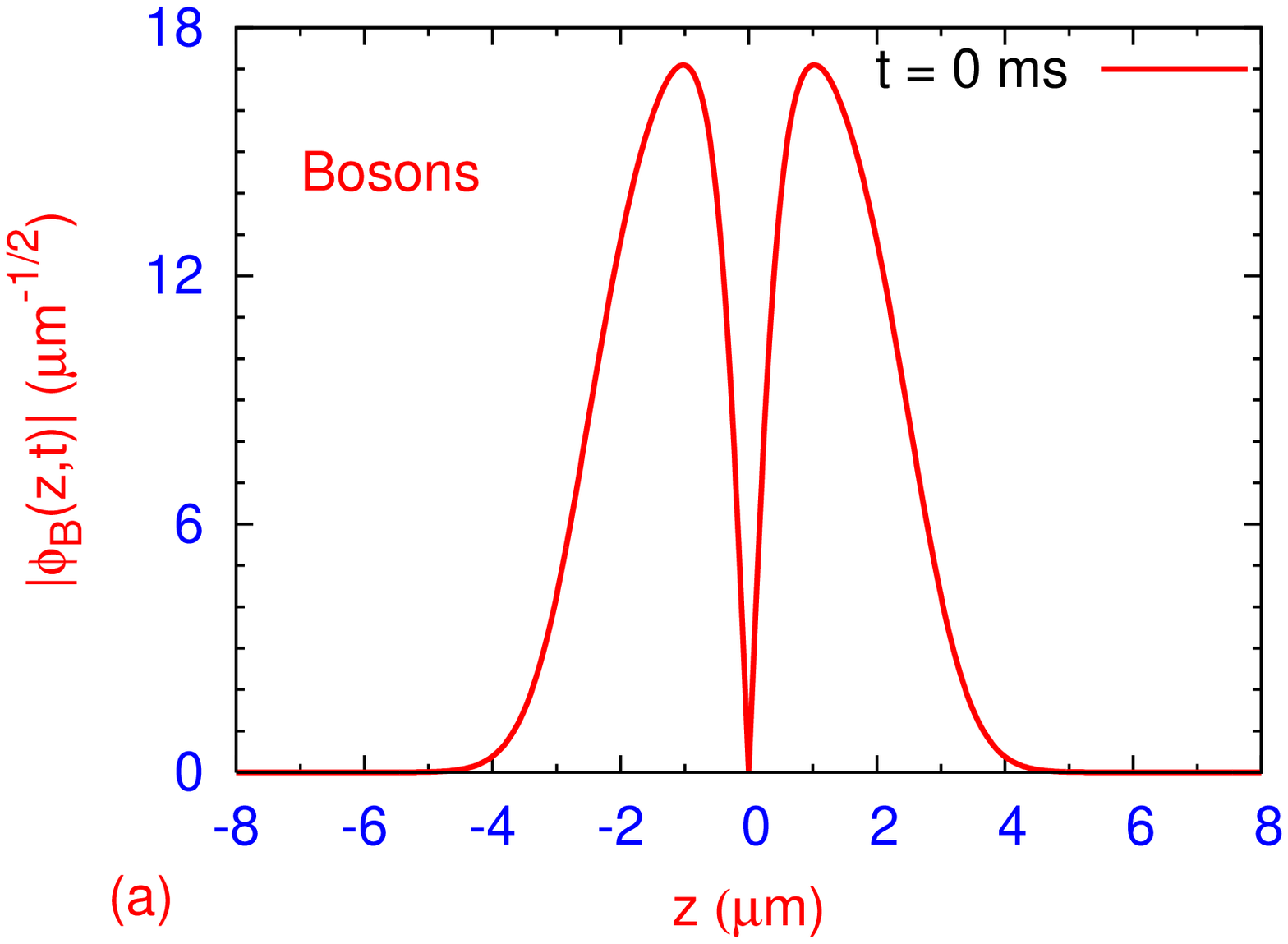}
\includegraphics[width=0.6\linewidth]{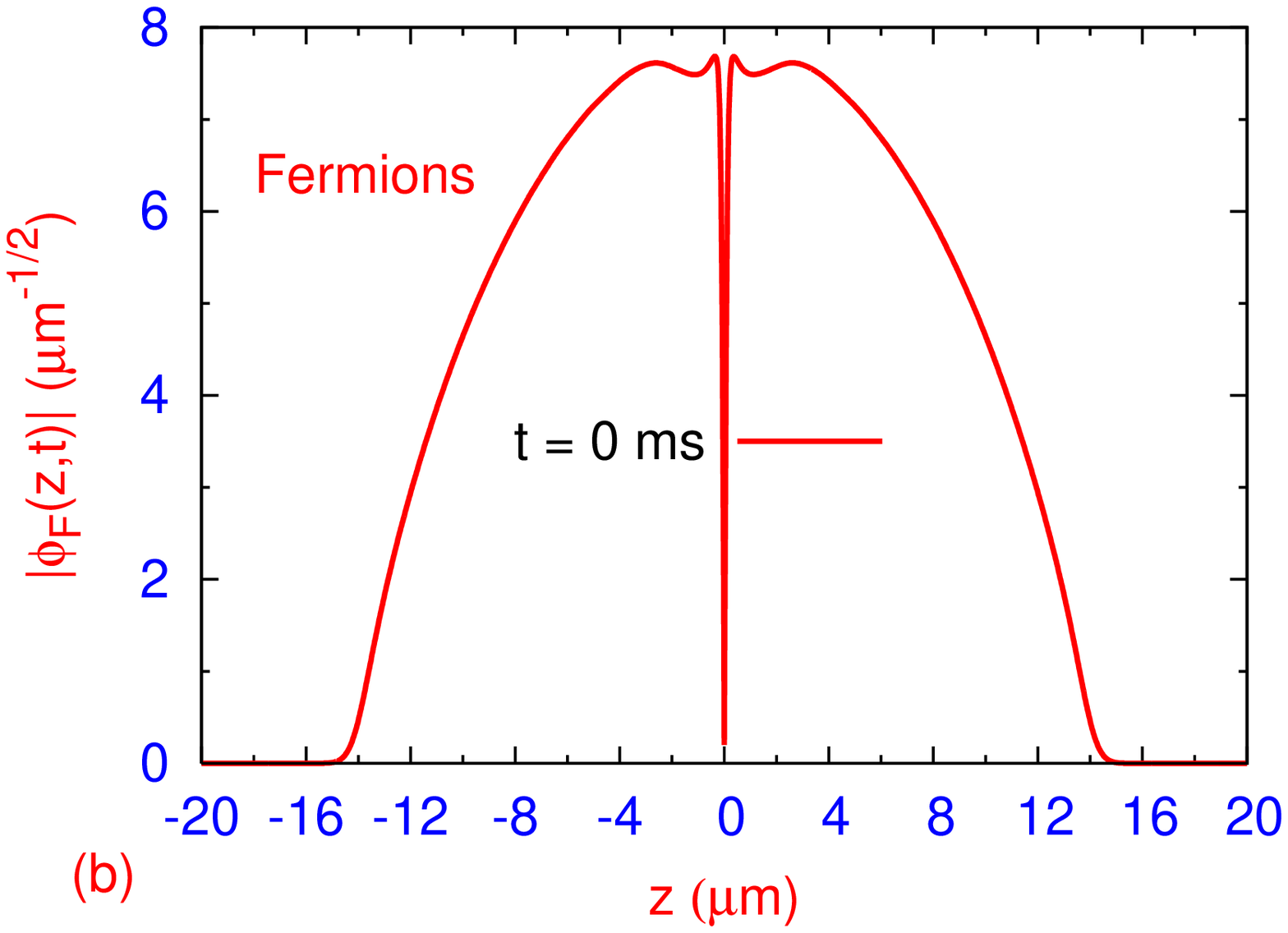}
\end{center}

\caption{The profiles of the oscillating function 
 $|\phi(z,t)|$ for (a) bosonic and
(b) fermionic dark solitons  of figure 1 (a)  vs. $z$.  
The oscillation was originated by jumping the boson-fermion scattering
length so that the nonlinearities $N_{BF}$ and $N_{FB}$ are suddenly
jumped from 40 to 44.  }
\end{figure}

\begin{figure}
 
\begin{center}
\includegraphics[width=0.8\linewidth]{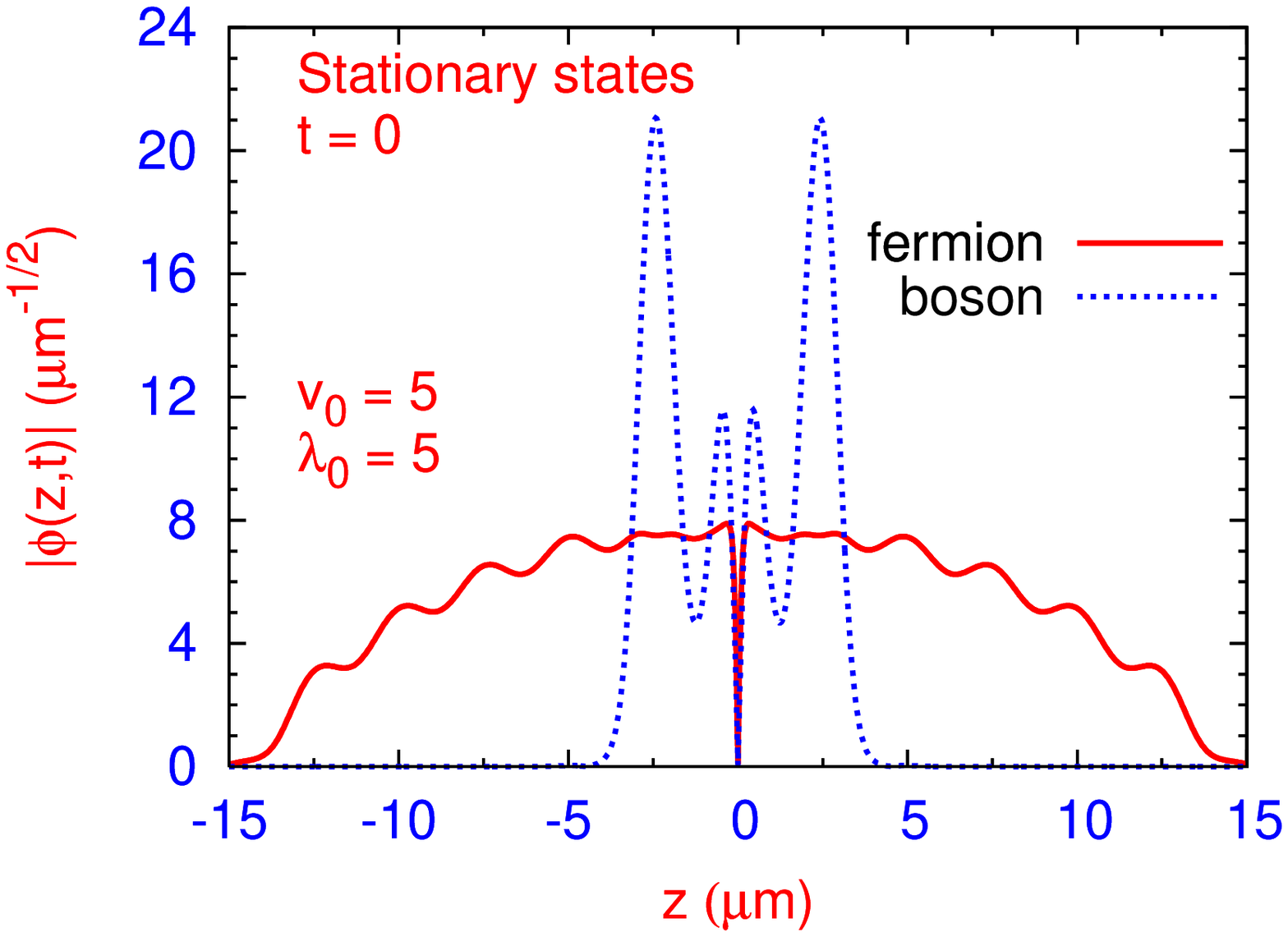}
\end{center}

\caption{ The
stationary function $|\phi(z,t)|$ for bosonic and
fermionic dark solitons  vs. $z$ with  $N_F=N_B=100$, $\nu=0.1$
 and $a_{BB}=a_{BF}=5 $ nm in the presence of an optical-lattice potential
with $\lambda_0 = 5$ and   
$v_0 = 5$.
 The nonlinearities are $N_{BB}=20, N_{BF}=40, N_{FB}=40,
N_{FF}=1275.$}

\end{figure}

The results of the present study on dark solitons are presented in
figures 1. In figure 1 (a) we plot the stationary functions $|\phi(z,t)|$
of
the boson
and fermion dark solitons. Both  functions have a notch in the
middle.  The notch in
the bosonic  function is wider than that in the fermionic 
function. In this case the fermionic nonlinearity (1275) is large. 
Consequently, the extension of the fermionic
function in figure 1 (a) is also much larger than the bosonic 
function.  

One way to observe the dark solitons experimentally is to allow
them to expand freely while the central notch will become wider in size to
be visible and photographed.  With this in mind  we study the free
expansion of the boson-fermion
mixture. The snapshots of the bosonic and fermionic  functions at regular
intervals of time during this expansion are shown in figures 1 (b) and
(c),
respectively. After expansion, the central notch in the bosonic dark
soliton expands as we find from figure 1 (b) but the
notch in the fermionic dark soliton in figure 1 (c) does not expand enough
to be
visible.  This will
make the fermionic dark soliton  more difficult to observe 
experimentally. However, this behavior is quite expected in a freely expanding
fermionic dark soliton  with a large nonlinear repulsion. In the fermionic
equation both the fermion-fermion and fermion-boson interactions are
highly repulsive.  The dark soliton  has a notch (hollow region) at the
center.
Because of the very strong repulsion, the DFG
tends to expand in all directions including the radially inward
direction to fill the central hollow space
as well as  outward directions. This inward repulsive force balances
partially the
outwards kinetic pressure and does not allow the central notch to expand
substantially during free expansion so as to be easily
observable. This is not the case for a moderately repulsive or
attractive pure single-component BEC, where the outward kinetic pressure
overcomes any repulsion among the atoms and 
the BEC  expands
only in
the radially outward direction with a widening of the notch. 
In the bosonic wave function of figure 1 (b)
the notch expands reasonably during the expansion of the BEC. However,
during the expansion in figures 1 (b) and (c) the central notch of both
the
bosonic and fermionic condensates 
has always a
zero at the origin.

Next we study the stability of the solitons illustrated in figures 1 under
a small perturbation inflicted by a sudden change in the boson-fermion
scattering length $a_{BF}$. After the solitons of figures 1 are formed we
increase
$a_{BF}$ by 10$\%$ at $t=0$ so that the nonlinearities $N_{BF}$ and
$N_{FB}$ are suddenly increased from 40 to 44. This can be performed
experimentally by varying a background magnetic field near a Feshbach
resonance in the boson-fermion system \cite{fesh}. The solitons then
execute small breathing oscillation around a mean position. The snapshot
of the soliton profiles shown in figures 3 (a) and (b) under this
perturbation demonstrates their stability.

Finally, we calculate the dark solitons in the boson-fermion mixture with
the parameters of figure 1 (a) in the presence of an optical-lattice
potential with $v_0=\lambda_0 = 5.$  
We plot in figure 3 the function $\phi(z,t)$
for the dark soliton in this case. 
The presence of the optical-lattice
potential creates modulations in both fermionic and bosonic 
functions. 
Because of the very strong
repulsion,  the fermions tend to occupy the
whole
available  region
in space and do not allow the formation of  pronounced notches with
hollow region inside at each optical-lattice site. Consequently, the
modulation in the fermionic 
function  is less pronounced  than the  modulation in the bosonic 
function.     The
bosonic function of figure 3 has pronounced notches compared 
to the smooth function  in figure 1 (a),  whereas the fermionic 
function of  figure 3 with small modulations is qualitatively more similar
to the fermionic  component  of  figure 1 (a).
We also studied free expansion of the dark solitons
in this case. The central notch in the fermionic dark soliton does not
also expand in this case. No new interesting physics emerges and the
results are not shown here.

\section{Summary}
 
We use  a coupled set of time-dependent mean-field-hydrodynamic
equations for a trapped degenerate boson-fermion mixture to study the
formation of a 
fermionic  dark soliton (vibrational excitation) 
in a DFG as
stationary
states. 
We calculate the stationary  functions with a notch at the center for
fermionic dark soliton 
of the
boson-fermion mixture. The existence of a central   notch in the 
wave function for a  dark soliton 
is typical of  vibrational  excitation. We perform numerical simulation of
the dark solitons for a  harmonic as well as a harmonic plus
optical-lattice traps.
The simulation is started with a time evolution of the mean-field
equations with the eigenfunction of the lowest excited state of the linear
oscillator problem setting all the nonlinearities to zero. The
nonlinearities are introduced slowly during time evolution and 
the iteration continued until convergence is obtained.  
We demonstrate the stability of the dark soliton upon the application of
a perturbation while the soliton executes small breathing oscillation.

The present time-dependent formulation also permits us to study
non-equilibrium free expansion of the coupled degenerate boson-fermion
mixture. One
way to observe the notch experimentally is to allow the  dark
soliton to undergo free expansion. 
We find that for a repulsive boson-fermion
interaction,
after a free expansion the notch in
the fermion function for a relatively small  fermion number 
of 100 does
not increase in
size significantly so as to be easily observable.  

In the present investigation  we used a set of mean-field equations for
the DFG-BEC mixture. A proper treatment of the DFG should be performed
using a fully antisymmetrized many-body Slater determinant wave
function \cite{yyy1}. Although
we believe the present conclusion about fermionic dark soliton to be true
in
general, it would be of interest to establish their existence using such a
fully antisymmetrized fermionic many-body wave function in the future.
Also, a  dark soliton features a particular spatial phase distribution,
i.e. a
step of $\pi$ phase. In a BEC, this phase profile is supported due to the
macroscopic phase of the condensate. Although a macroscopic phase of the
fermionic component should emerge in the present mean-field model, 
to the best of our knowledge
its
existence in a fermionic many-body wave function has not been
established rigorously. However, such a study is beyond the scope of this
paper and
would be a work of future interest. Nevertheless, it would be proper 
to call the fermionic excitation of the present paper by the term dark
soliton due to its appropriate density distribution with a central notch.

\ack 
 

The work is supported in part by the CNPq 
of Brazil.

\end{document}